\documentclass[preprint,showpacs,preprintnumbers,amsmath,amssymb,nofootinbib,natbib]{revtex4}

\usepackage{graphicx}% Include figure files
\usepackage{epsfig}		
\usepackage{dcolumn}% Align table columns on decimal point
\usepackage{bm}% bold math
\usepackage{amsmath}

\def\0{\mbox{\tiny $0$}}
\def\1{\mbox{\tiny $1$}}
\def\2{\mbox{\tiny $2$}}
\def\3{\mbox{\tiny $3$}}
\def\4{\mbox{\tiny $4$}}
\def\5{\mbox{\tiny $5$}}
\def\6{\mbox{\tiny $6$}}
\def\7{\mbox{\tiny $7$}}
\def\8{\mbox{\tiny $8$}}
\def\9{\mbox{\tiny $9$}}

\def\f14{\mbox{\tiny $\frac{1}{4}$}}

\begin{document}

\title{Lattice-layer entanglement in Bernal-stacked bilayer graphene}

\author{Victor A. S. V. Bittencourt}
\email{vbittencourt@df.ufscar.br}
\author{Alex E. Bernardini}
\email{alexeb@ufscar.br}
\affiliation{Departamento de F\'{\i}sica, Universidade Federal de S\~ao Carlos, PO Box 676, 13565-905, S\~ao Carlos, SP, Brasil}

\begin{abstract}
The complete {\em lattice-layer} entanglement structure of Bernal stacked bilayer graphene is obtained for the quantum system described by a tight-binding Hamiltonian which includes mass and bias voltage terms.
Through a suitable correspondence with the {\em parity-spin} $SU(2)\otimes SU(2)$ structure of a Dirac Hamiltonian, when it brings up tensor and pseudovector external field interactions,
the {\em lattice-layer} degrees of freedom can be mapped into such a {\em parity-spin} two qubit basis which supports the interpretation of the bilayer graphene eigenstates as entangled ones in a {\em lattice-layer} basis.
The Dirac Hamiltonian mapping structure simply provides the tools for the manipulation of the corresponding eigenstates and eigenenergies of the Bernal stacked graphene quantum system.
The quantum correlational content is then quantified by means of quantum concurrence, in order to have the influence of mass and bias voltage terms quantified, and in order to identify the role of the trigonal warping of energy in the intrinsic entanglement.
Our results show that while the mass term actively suppresses the intrinsic quantum entanglement of bilayer eigenstates, the bias voltage term spreads the entanglement in the Brillouin zone around the Dirac points.
In addition, the interlayer coupling modifies the symmetry of the {\em lattice-layer} quantum concurrence around a given Dirac point. It produces some distortion on the quantum entanglement profile which follows the same pattern of the iso-energy line distortion in the Bernal-stacked bilayer graphene.
\end{abstract}

\date{\today}
\maketitle

\section{Introduction}

The physical properties of graphene have been explored in both theoretical and experimental scopes in recent decades \cite{graph01, graph02, graph03, graph04, graphteo}. The quite singular structure of graphene energy bands, with a linear low energy profile driven by a massless Dirac-like equation, brings up complex implications for its electronic properties \cite{graph01}.
As is well-known, when graphene is under the action of a magnetic field, modified Landau levels are formed, leading to an anomalous behavior of its conductance \cite{graph05, graph06, graph07}. In particular, bilayer graphene exhibits still more peculiar properties due to its  weak interlayer coupling which also includes specific geometric arrangements between its layers \cite{graph08, graph010, graph011, graph012}. Different from the monolayer linear energy band profile, bilayer graphene has hyperbolic energy band profile near the corners of the first Brillouin zone, resembling the energy dispersion for free massive fermions.

That general correspondence of mono and bilayer graphene properties with the Dirac equation structure has already been exploited in the investigation of several relativistic-like features, from the {\em zitterbewegung} effect \cite{graph013, graph014, AlexD} to the Klein paradox \cite{graph015, AlexA}.
In fact, mapping the relativistic Dirac quantum mechanics into controllable physical systems is not exclusively encompassed by graphene quantum systems \cite{DiracEmu01, DiracEmu02, DiracEmu03, DiracEmu04, DiracEmu05, AlexB, AlexC}. For instance, the engineering of the Jaynes-Cummings Hamiltonian through ion traps has allowed one to simulate the Dirac dynamics \cite{DiracEmu01, Trapped01, Trapped02, Trapped03} in order to reproduce relativistic quantum effects \cite{Trapped01} as they are driven by several classes of external Dirac-like potentials \cite{Trapped02, Trapped03, Trapped04}.

Given that the solutions of the Dirac equation are supported by a $SU(2) \otimes SU(2)$ group structure driven by two internal degrees of freedom (DoF's), the spin and the intrinsic parity \cite{SU201, SU202}, the {\em parity-spin} entanglement profile of free or interacting particle solutions of the Dirac equation can be straightforwardly obtained \cite{SU201}. The $SU(2) \otimes SU(2)$ representation of Dirac bi-spinors are driven by a Hamiltonian dynamics written in terms of two qubit operators, for which the corresponding system eigenvectors can be identified and quantified as entangled states \cite{SU201,MeuPRA}.
The spin-parity entanglement exhibited by Dirac equation solutions is an example of intrinsic, or intraparticle, entanglement. Different from entanglement between degrees of freedom associated with distinct particles (for example polarization entanglement between different photons), intrinsic entanglement is encoded in internal degrees of freedom of a single particle. For instance, in the framework of neutron interferometry, the spin of the particle and a quantum number associated with different possible paths between the source and the measurement apparatus can be entangled \cite{Neut01,Neut02}. Due to the ability to manipulate and measure such neutron states, spin-path entanglement has been measured \cite{Neut01} and used to investigate, for example, Bell's inequality \cite{Neut02}. Also, spin-path entanglement can be suitably transferred to interparticle entanglement \cite{Neut03}. Another example emerges in quantum optics, where single particle entanglement can be encoded by single photons through different degrees of freedom, such as polarization and orbital angular momentum \cite{photon01}, polarization and transverse spatial degree of freedom \cite{photon02}, and in interferometer experiments \cite{photon03}. Quantum information protocols were engineered to take advantage of such intraparticle entanglement in photon systems \cite{photon04, photon05}.
Pragmatically, this interface between relativistic quantum mechanics and quantum information theory has been proved to be useful in classifying and quantifying the informational content of Dirac-like structures \cite{SU202,MeuPRA}.

In such a context, the inclusion of global potentials in the Dirac dynamics affects the correlational content of bispinors \cite{SU202, Barrier}.
By considering their invariance properties under Poincar\'{e} transformations, external field contributions to the Dirac dynamics are classified according to their (pseudo)scalar, (pseudo)vector and (pseudo)tensor characteristics \cite{Thaller}. A full Dirac Hamiltonian including all the above mentioned external field contributions should read \cite{SU202, Thaller}
\begin{eqnarray}
\label{E04T}
\hat{H}  &=& A^0(x)\,\hat{I}_4+ \hat{\beta}[ m + \phi_S (x) ] + \hat{\bm{\alpha}} \cdot [\hat{\bm{p}} - \bm{A}(x)] + i \hat{\beta} \hat{\gamma}_5 \mu(x) - \hat{\gamma}_5 q(x) + \hat{\gamma}_5 \hat{\bm{\alpha}}\cdot\bm{W}(x) \nonumber \\
&+& i \hat{\bm{\gamma}} \cdot [ \chi_a \bm{B}(x) + \kappa_a\, \bm{E}(x)  \,] + \hat{\gamma}_5 \hat{\bm{\gamma}}\cdot[\kappa_a\, \bm{B}(x)  - \chi_a \bm{E}(x) \,],
\end{eqnarray}
with $\hbar = c = 1$, $\hat{\bm{\gamma}} = \hat{\beta} \hat{\bm{\alpha}}$, and (the chirality matrix) $\hat{\gamma}_5 = -i \hat{\alpha}_x \hat{\alpha}_y \hat{\alpha}_z$, where
 $\hat{\beta}$ and $\hat{\bm{\alpha}} = \{\hat{\alpha}_x, \, \hat{\alpha}_y, \, \hat{\alpha}_z \}$ are the Dirac matrices that satisfies the anti-commuting relations
$\{\hat{\alpha}_i, \hat{\alpha}_j\} = 2 \, \delta_{ij} \, \hat{I}_4$, and
$\{\hat{\alpha}_i, \hat{\beta}\} =0$, with $i,j = x,y,z$, and
$\hat{\beta}^2 = \hat{I}_4$ (where $\hat{I}_N$ denotes the $N$-dimensional identity operator)
such that one can consider a particular representation\footnote{Dirac matrices are exhibited through different representations interconnected by unitary transformations.} given by
\begin{equation}
\hat{\bm{\alpha}} = \hat{\sigma}_x \otimes \hat{\bm{\sigma}} \equiv  \left[ \begin{array}{rr} 0 & \hat{\bm{\sigma}} \\ \hat{\bm{\sigma}} & 0 \end{array}\right],
\qquad \mbox{and} \qquad
\hat{\beta} = \hat{\sigma}_z \otimes \hat{I}_2 \equiv \left[ \begin{array}{rr} \hat{I}_2 & 0 \\ 0 & - \hat{I}_2 \end{array} \right],
\label{AAA}
\end{equation}
where, finally, ${\bm{\sigma}}$ are the Pauli matrices and, through out this paper, bold variables $\bm{a}$ denote vectors, with $a = \vert \bm{a} \vert = \sqrt{\bm{a} \cdot \bm{a}}$, and hats ``$~\hat{}~$'' denote operators.
The above Hamiltonian drives the dynamics of a fermion with mass $m$ and momentum $\bm{p}$ under an external vector field with time component $A^0(x)$ and spatial components $\bm{A} (x)$, nonminimally coupled to external magnetic and electric fields $\bm{B}(x)$ and $\bm{E}(x)$, respectively, through magnetic and electric moments, $\kappa_a$ and  $\chi_a$. This dynamics also includes an external pseudovector field with time component $q(x)$ and spatial components $\bm{W}(x)$, and both scalar and pseudoscalar fields, $\phi_S(x)$  and $\mu(x)$.

A generalized description of the effects of global potentials on the correlational content of Dirac bispinors was recently considered \cite{SU202}, and it can be specialized to some particular feasible physical systems. For instance, a trapped ion setup can be engineered to reproduced Dirac dynamics subject to global tensor and pseudo-tensor potentials, which describe a Dirac particle nonminimally coupled to external electric and magnetic fields \cite{Trapped03, MeuPRA}. The entanglement of Dirac equation solutions are then reinterpreted in terms of ionic state variables, and the intrinsic {\em parity-spin} entanglement is translated into the entanglement between quantum numbers related the total angular momentum and its projection onto the trapping magnetic field \cite{MeuPRA}. Moreover, entanglement shows a close relation to the average chirality of the state \cite{MeuPRA}.

Quantum entanglement has also been considered recently in the context of graphene physics, from the study of its relation with the quantum Hall effect \cite{CondMatter01, CondMatter02, CondMatter03}, for applications in quantum computing \cite{QuantumInfoGraph01, QuantumInfoGraph02, QuantumInfoGraph03, QuantumInfoGraph04}. In particular, when connected to the Hall effect, the entanglement spectrum \cite{CondMatter01} has a close relation to topological properties of condensed matter \cite{CondMatter04, CondMatter05}.
For quantum computing processes, the quantum entanglement is engendered either through spin-orbit couplings used to construct quantum gates between a graphene quantum dot and a flying qubit \cite{QuantumInfoGraph01}, or even by means of interactions between different valleys to process quantum information \cite{QuantumInfoGraph03, QuantumInfoGraph04}.

Such a fruitful scenario in the context of quantum information motivates one to find a relation between the intrinsic entanglement of Dirac equation solutions and the entanglement of bilayer graphene excitations. Considering bilayer graphene in its most stable configuration, the AB (or Bernal) stacking \cite{graph03, graph04}, it is shown that the tight binding Hamiltonian (including both bias voltage and mass terms) \cite{graph04, Predin}, when it is written in the reciprocal space, can be directly identified with a modified Dirac Hamiltonian including external pseudovector and tensor potentials.

In a straightfoward comparison to the trapped ion four-level system, where the framework involves a combination of tensor and pseudotensor Dirac potentials, according to the $SU(2)\otimes SU(2)$ Dirac structure for bilayer graphene the dynamics is engendered by pseudotensor and pseudovector Dirac potentials which, of course, produce different entangling properties.
Once a map between graphene and Dirac parameters is established, the single-particle excitations of bilayer graphene can be entirely described by the Dirac $SU(2) \otimes SU(2)$ structure.
By construction, the $SU(2) \otimes SU(2)$ {\em parity-spin} quantum correlational content is reinterpreted in terms of {\em lattice-layer} quantum entanglement that is intrinsic to the eigenstates of the graphene tight binding Hamiltonian. Such intrinsic lattice-layer entanglement is similar to other intraparticle quantum correlations observed in various physical systems, such as spin-path entanglement in neutron interferometry \cite{Neut01,Neut02} and single-photon entanglement \cite{photon01,photon02,photon03}.
Quantum entanglement will be quantified through the quantum concurrence in order to evince the effects of both bias voltage and mass terms on the entanglement of the AB tight binding eigenstates.
In parallel, the dynamical evolution of any graphene single excitation state can be suitably recovered and the effects of the trigonal warping, i.e. the distortion of iso-energy lines near Dirac points \cite{graph03,graph04, Predin} with respect to the above-mentioned intrinsic concurrence, can also be identified. Finally, it is worth mentioning that although the relation between bilayer graphene and relativistic quantum mechanics is well known, the explicit relation between the tight binding Hamiltonian and the Dirac Hamiltonian with external pseudotensor and pseudovector fields is derived, and it provides an effective and clean approach to evaluate analytically the eigensystem of the tight-binding Hamiltonian for the AB stacking and to recover the entangling properties of the fundamental state -- an approach that can be extended to other condensed matter systems, for instance, for the proposal of engendering quantum gates in order to implement quantum information protocols.

The paper is thus organized as follows. In Sec.~II the tight binding description of bilayer graphene in AB stacking is briefly reviewed and is connection with a modified Dirac Hamiltonian including some particular external potentials is fully identified. In Sec.~III the eigenstates are recovered by means of a suitable easily used ansatz, which takes into account specific algebraic properties of the Dirac Hamiltonian. The band structure of bilayer graphene and its properties are all obtained and the single excitation dynamics of a generic initial state is recovered. In Sec.~IV the entangling properties of the Bernal stacked graphene eigenstates are quantified by means of quantum concurrence, and the effects due to bias voltage and mass terms are analyzed. The final issue is concerned with the changes to quantum concurrence when it is affected by the trigonal warping.
Our final conclusions are drawn in Sec.~V, in order to point out a general overview of the entanglement properties of the Bernal-stacked bilayer graphene.

\section{AB tight binding Hamiltonian and the Dirac equation}

\begin{figure}[!b]
\begin{minipage}{0.4\textwidth}
\centering
\includegraphics[width= 1.1\linewidth]{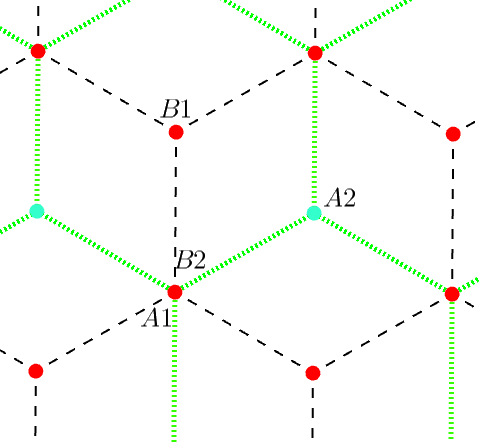}
\end{minipage}
\hspace{1 cm}
\begin{minipage}{0.4\textwidth}
\centering
\includegraphics[width= 1.1\linewidth]{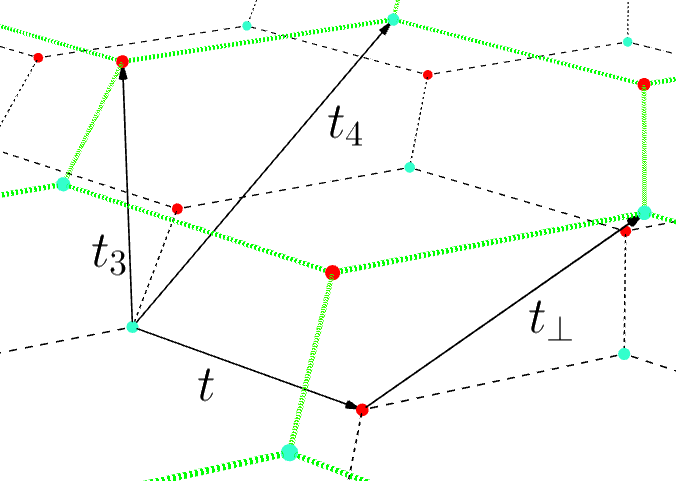}
\end{minipage}
\renewcommand{\baselinestretch}{1.0}
\caption{(Left) Top view of the geometrical configuration of the AB (Bernal) stacking. Half of the atoms of the upper layer (joined by dotted lines) are exactly above half of the atoms of the lower layer (joined by dashed lines). Sites that are placed exactly above a site of the lower layer are called dimer sites (A1 and B2), while sites that are localized above the center of the other honeycomb are called nondimer sites (B1 and A2). (Right) Schematic representation of the hopping amplitudes of the tight binding model for the bilayer graphene. The parameter $t$ describes the hopping between next neighbors in the same layer; $t_\bot$ is the hopping from a nondimer site to its nearest nondimer site; $t_3$ is the hopping from a dimer site to its nearest dimer site, and $t_4$ is the hopping from a dimer to the nearest nondimer site.}
\label{ABstacking}
\end{figure}

The geometry of the AB stacking (or Bernal stacking) consists of two layers of graphene arranged such that half of the atoms of the upper layer are localized above half of the atoms of the lower layer (dimer sites), while the other half atoms are localized above the center of the lower honeycombs (nondimer sites) \cite{graph03, graph04}, as depicted in Fig.~\ref{ABstacking}.
Labeling the sublattices of layer 1 as $A1$ and $B1$, and the sub-lattices of layer 2 as $A2$ and $B2$ (see Fig.~\ref{ABstacking}) the tight binding Hamiltonian is given by \cite{graph03,graph04,OldGraph01,OldGraph02}
\begin{eqnarray}
\label{tightbinding}
\hat{\mathcal{H}}_{AB} = &-& t \displaystyle \sum_{\bm{k}} \left[ \, \Gamma(\bm{k}) \hat{a}_{1\bm{k}} ^\dagger \hat{b}_{1\bm{k}} + \Gamma(\bm{k}) \hat{a}_{2 \bm{k}}^\dagger \hat{b}_{2 \bm{k}} + \mbox{H.c.} \,\right] \nonumber \\
 &+& t_{\bot} \displaystyle \sum_{\bm{k}} \left[ \, \hat{b}_{1 \bm{k}}^\dagger \hat{a}_{2 \bm{k}} + \hat{a}_{2 \bm{k}}^\dagger \hat{b}_{1 \bm{k}} \, \right] - t_3 \displaystyle \sum_{\bm{k}} \left[\, \Gamma(\bm{k}) \hat{b}_{2 \bm{k}}^\dagger \hat{a}_{1 \bm{k}} + \Gamma^* (\bm{k}) \hat{a}_{1 \bm{k}}^\dagger \hat{b}_{2 \bm{k}}  \, \right] \nonumber \\
 &+&t_4 \displaystyle \sum_{\bm{k}} \left[ \, \Gamma(\bm{k})(\hat{a}_{1 \bm{k}}^\dagger \hat{a}_{2 \bm{k}} + \hat{b}_{1 \bm{k}}^\dagger \hat{b}_{2 \bm{k}}) + \mbox{H.c.}  \, \right],
\end{eqnarray}
where $\hat{\alpha}_{i \bm{k}}^\dagger$ is the creation operator for an excitation on the $\alpha$ lattice in the $i$ layer with wave vector $\bm{k}$, and $\Gamma(\bm{k}) = \displaystyle \sum_{j=1} ^3 e^{ i \bm{k} \cdot \bm{\delta}_{j}}$ is given in terms of the next-neighbor vectors $$\bm{\delta}_{1,2} = \left( - a/2, \, \pm a \sqrt{3}/2 \right), \hspace{0.5 cm} \bm{\delta}_3 = \left(a, 0 \right).$$ A schematic representation of the hopping amplitudes $t$, $t_\bot$, $t_3$, and $t_4$ is also depicted in Fig.~\ref{ABstacking}.

The tight binding Hamiltonian (\ref{tightbinding}) is an effective description of the underlying dynamics of the bilayer graphene in the AB stacking which is often used to study its electronic and optical properties \cite{graph03, graph04}. In contrast to the usual energy dispersion relation predicted for the monolayer graphene, which is totally symmetric around the Dirac point, the tight binding model (\ref{tightbinding}) predicts a distortion of the iso-energy lines near the Dirac points, as a consequence of the inclusion of the hopping $t_3$, the so-called trigonal warping \cite{graph03, graph04, Predin, Trig01, Trig02}.

To derive analytical properties of the dynamics driven by (\ref{tightbinding}), one sets $t_4=0$. In the $\bm{k}$ space, the Hamiltonian $\hat{\mathcal{H}}_{AB}$ is then written in the basis $\{ \vert A1 (\bm{k}) \rangle,  \vert B1 (\bm{k}) \rangle, \vert A2 (\bm{k}) \rangle, \vert B2 (\bm{k}) \rangle \}$ ($\vert \alpha _i (\bm{k}) \rangle = \hat{\alpha}_{i \bm{k}} ^\dagger \vert 0 \rangle$), in order to stay as
\begin{equation}
\label{ABHamiltonian}
\hat{\mathcal{H}}_{AB} =\left[ \, \begin{array}{cccc}
0 & - t \Gamma(\bm{k}) & 0 & - t_3 \Gamma^*(\bm{k})\\
- t \Gamma^*(\bm{k}) & 0 & t_\bot & 0\\
0 & t_\bot & 0& -t\Gamma (\bm{k})\\
- t_3 \Gamma(\bm{k}) & 0 & - t \Gamma^*(\bm{k}) & 0
\end{array} \right].
\end{equation}
Two additional interactions might be added to the above dynamics, both associated with an energy gapping \cite{graph03, Predin}: a mass term, $\hat{\mathcal{H}}_m$, and a bias voltage term, $\hat{\mathcal{H}}_{\Lambda}$, given by
\begin{equation}
\label{massHamiltonian}
\hat{\mathcal{H}}_{m}= \mbox{diag}\{ m, \, -m, \, m, \, -m\},
\end{equation}
\begin{equation}
\label{biasvoltageHamiltonian}
\hat{\mathcal{H}}_{\Lambda}= \mbox{diag} \left\{ \frac{\Lambda}{2}, \, \frac{\Lambda}{2}, \, - \frac{\Lambda}{2}, \, - \frac{\Lambda}{2} \right\},
\end{equation}
such that the total Hamiltonian reads
\begin{eqnarray}
\label{Htotal}
\hat{\mathcal{H}} = \hat{\mathcal{H}}_{AB} + \hat{\mathcal{H}}_m + \hat{\mathcal{H}}_{\Lambda}.
\end{eqnarray}

The tight binding model for monolayer graphene predicts an energy dispersion relation which is approximately quadratic near the corners of the first Brillouin zone \cite{graph03, graph05, graph012}. Around this point, the electronic excitation of monolayer graphene behaves like massless Dirac fermions, and an effective description in terms of Dirac equation in a reduced $2+1$ dimension can be used to derive its physical properties \cite{graph03, graph04, graph05, graph08, graph010}. For the bilayer graphene, the AB tight binding Hamiltonian including both mass and bias voltage (\ref{Htotal}) reproduces the dynamics of a modified Dirac equation on the entire $\bm{k}$ space. Like the Jaynes-Cummings Hamiltonian -- which was written as a Dirac-like Hamiltonian in terms of ion trap parameters \cite{DiracEmu01, Trapped01, Trapped02, Trapped03, MeuPRA} -- the Hamiltonian from (\ref{Htotal}) can also be mapped into a Dirac Hamiltonian in momentum space. Nevertheless, the combinations of external fields arising from such mapping are different from those described in the trapped ion system, which is engineered to reproduce the Dirac equation and is not intrinsic to the system dynamics.
To derive such a relation, one notices that (see Appendix for matrix manipulations), $\hat{\mathcal{H}}$ can be rewritten as
\begin{eqnarray}
\hat{\mathcal{H}} &=&
\frac{t_\bot}{2}\left( \hat{\alpha}_x - i \hat{\gamma}_y\right)
- t \left\{\mbox{Re}[\Gamma (\bm{k})] \hat{\gamma}_5 \hat{\alpha}_x -\mbox{Im}[\Gamma(\bm{k})] \hat{\gamma}_5 \hat{\alpha}_y\right\}\nonumber\\
&&\qquad\qquad -\frac{t_3}{2}\left\{ \mbox{Re}[\Gamma (\bm{k})](\hat{\alpha}_x + i\,\hat{\gamma}_y) + \mbox{Im}[\Gamma(\bm{k})]( \hat{\alpha}_y - i\, \hat{\gamma}_x)\right\}
+ m \hat{\gamma}_5 \hat{\alpha}_z
+\frac{\Lambda}{2}\hat{\beta},
\end{eqnarray}
which allows one to map the total Hamiltonian $\hat{\mathcal{H}}$ into the modified Dirac Hamiltonian (in momentum space) in order to have
\begin{equation}
\label{DiracHamiltonian}
\hat{\mathcal{H}} = \bm{p} \cdot \hat{\bm{\alpha}} + M \hat{\beta} + \bm{W} \cdot \hat{\gamma}_5 \hat{\bm{\alpha}} + i \bm{\mathcal{E}} \cdot \hat{\bm{\gamma}},
\end{equation}
which contains the usual free massive particle term, $\bm{p} \cdot \hat{\bm{\alpha}} + M \hat{\beta}$, plus pseudovector and tensor potential contributions, $\bm{W} \cdot \hat{\gamma}_5 \hat{\bm{\alpha}}$ and $i \bm{\mathcal{E}} \cdot \hat{\bm{\gamma}}$. By comparing Eqs. (\ref{rel01}) - (\ref{rel03}) with Eq.~(\ref{DiracHamiltonian}) one identifies the following one-to-one correspondence between graphene and Dirac parameters:
\begin{eqnarray}
\label{relations}
\bm{p} &\leftrightarrow& \frac{t_\bot - t_3 \mbox{Re}[\Gamma (\bm{k})]}{2} \bm{i} - \frac{t_3 \mbox{Im}[\Gamma(\bm{k})]}{2} \bm{j}, \hspace{1.0 cm } M \leftrightarrow \frac{\Lambda}{2} ,\nonumber \\
\bm{W} &\leftrightarrow& - t \mbox{Re}[\Gamma (\bm{k})] \bm{i} + t \mbox{Im}[\Gamma(\bm{k})] \bm{j} + m \bm{l}, \hspace{1 cm} \bm{\mathcal{E}} \leftrightarrow \frac{ t_3 \mbox{Im}[\Gamma(\bm{k})]}{2} \bm{i} - \frac{t_\bot + t_3 \mbox{Re}[\Gamma(\bm{k})]}{2} \bm{j},
\end{eqnarray}
where $\{\bm{i}, \bm{j} , \bm{l} \}$ are the unitary vectors. In this framework, the tight binding Hamiltonian (\ref{Htotal}) simulates the Dirac Hamiltonian with the external potentials as they appear in Eq.~(\ref{DiracHamiltonian}). Therefore, Dirac Hamiltonian eigenstates, $\vert \psi_{n,s} \rangle$ $(n,s = \{0,1\})$, are encoded by superpositions of the graphene one particle states given by
\begin{equation}
\label{eigenstates}
\vert \psi_{n,s} (\bm{k}) \rangle \equiv M^{A1}_{n,s}\, \,\vert A1 (\bm{k})  \rangle + M^{B1}_{n,s} \, \,\vert B1 (\bm{k})  \rangle + M^{A2}_{n,s}  \,\,\vert A2 (\bm{k})  \rangle+ M^{B2}_{n,s} \, \,\vert B2 (\bm{k})  \rangle.
\end{equation}

The Dirac Hamiltonian eigenstates, $\vert \psi_{n,s} \rangle$, are supported by a $SU(2) \otimes SU(2)$ group structure involving their internal DoF's (assigned to a Hamiltonian dynamics) written in terms of the direct product of Pauli operators \cite{SU201, SU202}. In fact, according to the  representation of Dirac matrices adopted here [c. f. Eq.~(\ref{AAA})], the Hamiltonian (\ref{DiracHamiltonian}) reads $$\hat{\mathcal{H}} =  \bm{p} \cdot ( \hat{\sigma}_x ^{(1)} \otimes \hat{\bm{\sigma}}^{(2)}) + M (\hat{\sigma}_z^{(1)} \otimes \hat{I}^{(2)}) + \bm{W} \cdot ( \hat{I}^{(1)} \otimes \hat{\bm{\sigma}}^{(2)}) -  \bm{\mathcal{E}} \cdot ( \hat{\sigma}_y^{(1)} \otimes \hat{\bm{\sigma}}^{(2)}), $$ which drives the dynamics of two discrete DoF's labeled by $(1)$ and $(2)$. These DoF's are associated to a system $\mathcal{S}$ composed of two sybsystems, $\mathcal{S}_1$ (associated with the spin DoF) and $\mathcal{S}_2$ (associated to with intrinsic parity DoF), described by a composite Hilbert space $H = H_1 \otimes H_2$ with $\mbox{dim}H_1 = \mbox{dim}H_2 = 2 $ \cite{SU201, SU202}.
Therefore, the corresponding bispinors, which are bipartite states in this framework, are {\em parity-spin} entangled states, and the above structure establishes the condition for the computation of separability quantifiers \cite{Entanglement01}.

The one-to-one relation between the total bilayer graphene Hamiltonian $\hat{\mathcal{H}}_{AB}$ and the modified Dirac Hamiltonian (\ref{DiracHamiltonian}) sets a correspondence between the discrete DoF's of the Dirac equation and those intrinsic ones of the bilayer graphene. One can identify the bilayer graphene DoF's as lattice ($A$ or $B$) and layer ($1$ or $2$) DoF's, such that the {\em parity-spin} entanglement intrinsic to solutions of the Dirac equation can be translated into a {\em lattice-layer} entanglement. Therefore all states of bilayer graphene can be interpreted as two-qubit states through the two-qubit assignment
\begin{eqnarray}
\label{assi}
\vert A1 \rangle &\equiv& \vert 00 \rangle, \hspace{0.5 cm} \vert B1 \rangle \equiv \vert 01 \rangle, \nonumber \\
\vert A2 \rangle &\equiv& \vert 10 \rangle, \hspace{0.5 cm} \vert B2 \rangle \equiv \vert 11 \rangle,
\end{eqnarray}
adopted from now on.

\section{Eigenstates of the Dirac equation and its correspondence to the eigenstates of the AB tight binding Hamiltonian}

As mentioned above, the invariance of Dirac equation under Poincar\'{e} transformations provides a systematic classification of external potentials according to their scalar, pseudoscalar, vector, pseudovector, tensor, and pseudotensor transformation properties \cite{Thaller}. For specific combinations of these potential terms, the corresponding modified Dirac Hamiltonian exhibits algebraic properties which can be resumed by an ansatz procedure for computing its eigenstates \cite{SU202,MeuPRA, New}. Since the Hamiltonian from (\ref{DiracHamiltonian}) includes only tensor and pseudovector additional terms -- with respect to the free particle preliminary content -- the density matrices associated to each one of the eigenstates can be easily obtained \cite{SU202,MeuPRA}.
From the properties of the Dirac matrices, one firstly notices that
\begin{eqnarray}
\label{prop01}
\hat{\mathcal{H}}^2 &=& g_1 \hat{I}_4 + 2 \hat{\mathcal{O}},
\end{eqnarray}
where $\hat{\mathcal{O}}$ is a traceless operator given by
\begin{equation}
\hat{\mathcal{O}} = (\bm{p} \cdot \bm{W}) \hat{\gamma}_5 + i (\bm{W}\cdot \bm{\mathcal{E}}) \hat{\beta} \hat{\gamma}_5 - [\, M \bm{W} + (\bm{p} \times \bm{\mathcal{E}}) \,] \cdot \hat{\gamma}_5 \hat{\bm{\gamma}},
\end{equation}
which satisfies
\begin{eqnarray}
\label{prop02}
\hat{\mathcal{O}}^2 = \frac{1}{4}(\hat{\mathcal{H}}^2 - g_1 \hat{I}_4)^2 = g_2 \hat{I}_4,
\end{eqnarray}
with
\begin{eqnarray}
g_1 &=& \frac{1}{4}\mbox{Tr}[\hat{\mathcal{H}}^2] = p^2 + M^2 + W^2 + \mathcal{E}^2,
\end{eqnarray}
and
\begin{eqnarray}
g_2 &=& \frac{1}{16}\mbox{Tr}\left[\,( \hat{\mathcal{H}}^2 - \frac{1}{4}\mbox{Tr}[\hat{\mathcal{H}}^2] )^2 \right] = \nonumber \\
&=& M^2 W^2 + 2 M \bm{W} \cdot(\bm{p} \times \bm{\mathcal{E}}) + \vert \bm{p} \times \bm{\mathcal{E}} \vert^2 + (\bm{p} \cdot \bm{W})^2 + (\bm{W} \cdot \bm{\mathcal{E}})^2.
\end{eqnarray}

For the Hamiltonian satisfying Eq.~(\ref{prop01}), the density matrices of pure eigenstates $\rho_{n,s} = \vert \psi_{n,s} \rangle \langle \psi_{n,s} \vert$ are given by \cite{SU202}
\begin{equation}
\label{ansatz}
\rho_{n,s} = \frac{1}{4}\left[\hat{I}_4 + \frac{(-1)^n}{\vert \lambda_{n,s} \vert} \, \hat{\mathcal{H}} \right] \left[ \hat{I}_4 + \frac{(-1)^s}{\sqrt{g_2}} \, \hat{\mathcal{O}} \right],
\end{equation}
which are indeed stationary states of the corresponding Liouville equation $[ \rho_{n,s}, \hat{\mathcal{H}}] = 0$. The averaged energies of the states (\ref{ansatz}) correspond to the eigenvalues $\lambda_{n,s}$ associated to $\rho_{n,s}$, which is evaluated by $(\lambda_{n,s} - g_1)^2 = 4g_2$, i. e.,
\begin{eqnarray}
\label{eigenvalues0}
\lambda_{n,s} = \mbox{Tr}[\hat{\mathcal{H}} \rho_{n,s}] = (-1)^n\sqrt{g_1 + 2 (-1)^s \sqrt{g_2}}.
\end{eqnarray}
Through this procedure, one recovers the full single-particle energy spectrum in $\bm{k}$ space by means of (\ref{relations}), such that the energy eigenvalues (\ref{eigenvalues0}) read [for $\Gamma(\bm{k}) = \vert \Gamma(\bm{k}) \vert e^{i \, \phi(\bm{k})}$]
\begin{eqnarray}
\label{eigenvalues}
\lambda_{n,s}(\bm{k}) &=& (-1)^n \bigg[ \frac{1}{2} \bigg(  2 t^2 \vert \Gamma(\bm{k}) \vert^2 + t_\bot^2 + t_3 \vert \Gamma(\bm{k}) \vert^2 + 2 m^2 + \frac{\Lambda^2}{2} \nonumber \\
 &&\qquad\qquad\qquad+ (-1)^s [4 t^2\, \vert \Gamma(\bm{k}) \vert^2 (\,t_\bot^2 + \Lambda^2+ t_3^2 \vert \Gamma(\bm{k}) \vert^2 - 2 t_\bot t_3 \cos(3 \phi(\bm{k}))) \nonumber \\ &&\qquad\qquad\qquad\qquad\qquad\qquad + ( t_3 \vert \Gamma(\bm{k}) \vert^2 - t_\bot^2 + 2 m \Lambda )^2  ]^{1/2} \, \bigg) \bigg]^{1/2}.
\end{eqnarray}

\begin{figure}[!htb]
\includegraphics[width= 0.75\linewidth]{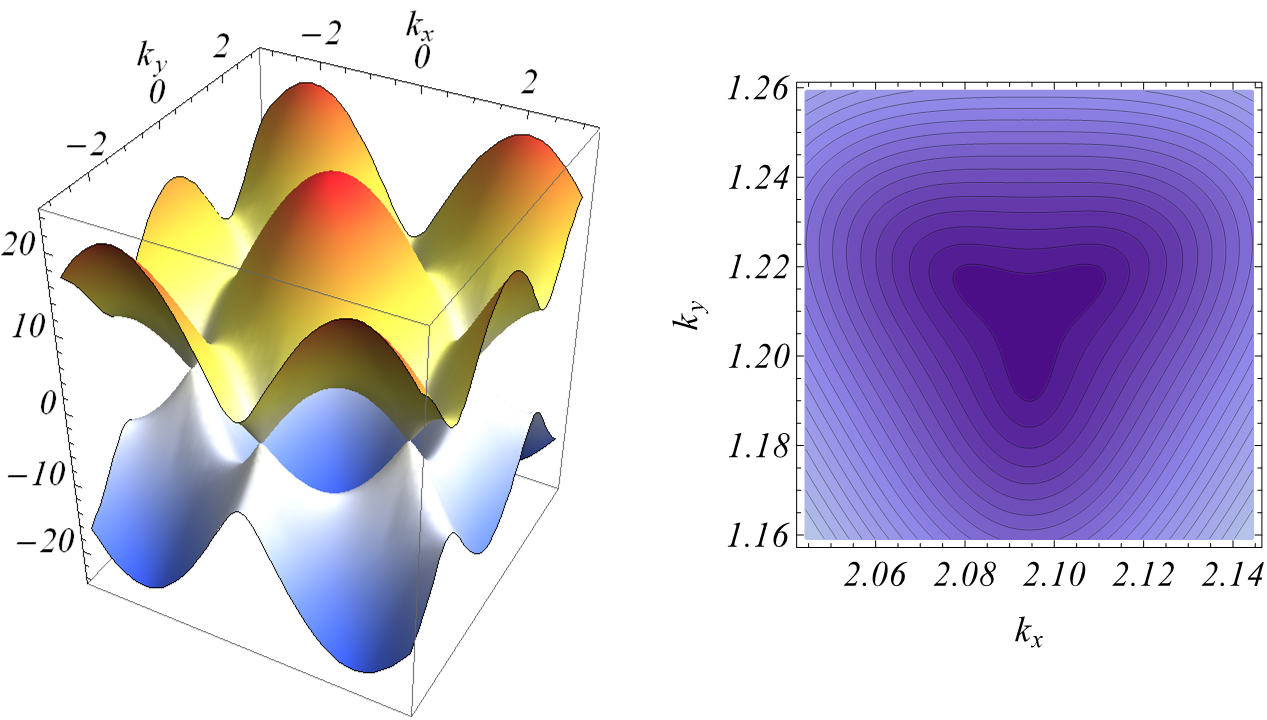}
\renewcommand{\baselinestretch}{1.0}
\caption{(Left) Energy bands for the branch $s = 1$, $t/t_\bot = 8.29$, $t_3/t_\bot = 0.99$, and for $m/t_\bot = \Lambda/t_\bot = 0$. The energy bands touch the corners of the first Brillouin zone (Dirac points). Near such points the energy profile approaches a parabola in $k$ and the dispersion relation reproduces a massive Dirac equation. (Right) Zoom around the Dirac point. The isoenergy lines are distorted due to the $t_3$ hopping -- the trigonal warping effect. The constant energy lines have a $\frac{2 \pi}{3}$ symmetry around these points due to the $\cos(\,3 \phi(\bm{k})\,)$ term in (\ref{eigenvalues}). In the absence of the $t_3$ hopping, the iso-energy lines around a Dirac point are perfectly symmetric.}
\label{energy}
\end{figure}

\begin{figure}[!htb]
\includegraphics[width= 8 cm]{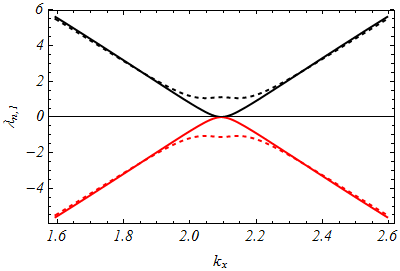}
\renewcommand{\baselinestretch}{1.0}
\caption{Auxiliary plot for the energy bands for the branch $s=1$ as function of $k_x$ for $k_y = \frac{2 \pi}{3 \sqrt{3} a} $, $t/t_\bot = 8.29$, $t_3/t_\bot = 0.99$, and for $m/t_\bot = \Lambda/t_\bot = 0$ (solid lines) and $m/t_\bot = \Lambda/t_\bot = 1$ (dashed lines). For $m = \Lambda = 0$ the bands $n = 0$ and $n = 1$ touch the Dirac point. Non-zero values of $m$ and $\Lambda$ open an energy gap between the valence and conduction band, deforming the linear dispersion (for $m=\Lambda = 0$) around the Dirac point into a hyperbolic one.}
\label{gap}
\end{figure}

The two inequivalent values of $s$ ($0$ and $1$) define two energy branches composed by two energy bands, corresponding to $n=0$ and $n=1$. The energy eigenvalues (\ref{eigenvalues}) exhibit extremum points for specific values of the wave vector $\bm{k}$, as depicted in the left plot of Fig.~\ref{energy} for the energy branch $s = 1$, $t/t_\bot = 8.29$, and $t_3/t_\bot = 0.99$, which are in correspondence with the experimental measurements from Ref.~\cite{Exp01} (in the case of $m/t_\bot = \Lambda/t_\bot = 0$). Two extremum energy points occur for $\Gamma(\bm{k}) = 0$, which correspond to the two inequivalent Dirac points $\bm{K}_\pm = \frac{2 \pi}{3 \sqrt{3} a} \left( \sqrt{3}, \pm 1 \right)$. Around the Dirac points, the constant energy lines are distorted [see the right (zoom) plot of Fig.~\ref{energy}], and for larger values of the parameter $t_3/t_\bot$ additional local minimum points, given by the condition $\cos{(3 \phi(\bm{k}))} = 0$ and $\vert \Gamma (\bm{k}) \vert = \frac{t_\bot t_3}{t^2}$, are identified. The appearance of these orbiting the Dirac points is a consequence of the $t_3$ hopping in $\hat{\mathcal{H}}_{AB}$, which produces a distortion effect of the isoenergy lines -- the so-called trigonal warping \cite{graph03, graph04}. For $m = \Lambda = 0$, Dirac points correspond to contact points between upper and lower energy bands, where nonzero values of those parameters produce an energy gap between such bands, as depicted in Fig.~\ref{gap}. The branch $s = 0$ always exhibits an energy gap between its energy bands. Qualitatively, its profile is the same as that of the branch $s=1$.

The eigenstates and eigenenergies of the tight binding Hamiltonian for bilayer graphene were previously evaluated through both algebraic techniques \cite{graph01,graph03}\footnote{See also the appendix of \cite{Predin} for a complete derivation of the eigensystem for bilayer graphene.} and numerical procedures, which can incorporate for instance effects of impurities \cite{Yuan01}. Nevertheless, the approach adopted here for bilayer graphene is self-consistent once it demands less algebraic work and generates analytical results based on previously constructed solutions of the Dirac equation with external fields \cite{SU202}. Moreover, the relation with Dirac dynamics elucidates the nature of quantum correlations in the eigenstates of $\hat{\mathcal{H}}$ as arising from a $SU(2)\otimes SU(2)$ structure of the Hamiltonian. One may also extend the method for construction of the density matrices associated with the eigenstates to any Hamiltonian $\hat{H}$ satisfying the conditions $\hat{H}^2 = c_1 \hat{I} + 2 \hat{O}$ and $\hat{O}^2 = c_2 \hat{I}$, and it is possible to directly include suitable open system effects by means of the Kraus operator formalism, following a procedure similar to the one presented in \cite{New}.

\section{Entanglement properties of the AB eigenstates}

Once the eigenstates of the tight binding Hamiltonian are recovered by the ansatz Eq.~(\ref{ansatz}), it is possible to construct the dynamical evolution of any initial state $\rho(0)$. By using the eigenstate completeness relation, $\sum_{(n,s) = 0}^1 \rho_{n,s} = 1$, one has
\begin{equation}
\rho (t)  = e^{- i \hat{H}_D t} \rho(0)  e^{ i \hat{H}_D t} = \displaystyle \sum^1_{n,s=0}\sum^1_{m,l=0} e^{- i (\lambda_{n,s} - \lambda_{m,l}) t}\, \varrho_{n,s} \, \rho (0) \, \varrho_{m,l},
\end{equation}
which allows one to compute any physical observable $\hat{A}$ through $\langle \hat{A} \rangle (t) = \mbox{Tr}[\hat{A} \, \rho(t)]$.
The eigenstates $\rho_{n,s}$ reflect the $SU(2) \otimes SU(2)$ structure of the Hamiltonian $\hat{\mathcal{H}}$, which allows one to describe the systems driven by its dynamics as a composite quantum systems with two discrete DoF's \cite{SU202, MeuPRA, Barrier}.
In this case, a bipartite state described by a density matrix $\rho \in H = H_1 \otimes H_2$, such as those constructed by means of (\ref{ansatz}), is separable if \cite{Entanglement01}
\begin{equation}
\rho = \displaystyle \sum_i w_i \,\varrho_i^{(1)} \otimes \varrho_i^{(2)},
\end{equation}
where $\varrho^{(1)}_i \in H_1$, $\varrho^{(2)}_i \in H_2$, and $\displaystyle \sum w_i = 1$.

In fact, different quantities can be used as entanglement quantifiers. For instance, entanglement entropy defined as the von Neumann entropy of the reduced state $\rho_{1,2} = \mbox{Tr}_{2,1}[\rho]$:
\begin{equation}
E_{\mbox{vN}} [ \rho ] = S[\rho_2] = - \mbox{Tr}_2[\rho_2 \mbox{log}_2 \rho] = S[\rho_1] = - \mbox{Tr}_1[\rho_1 \mbox{log}_1 \rho],
\end{equation}
is an entanglement quantifier for pure states \cite{Entanglement02, Entanglement03}. The above equality is guaranteed by the fact that, for pure states $\rho$, the reduced density matrices $\rho_{1,2} = \mbox{Tr}_{2,1}[\rho]$ have identical eigenvalues and, if the state is entangled, then either $\rho_{1(2)}$ are mixed (the Schmidt theorem) \cite{Entanglement02}. Likewise, the quantum concurrence $\mathcal{C}[\rho]$ -- whose definition is primarily related to the entanglement of formation \cite{Entanglement04} -- is an entanglement quantifier more convenient for the proposal of this work. For any state $\rho$, the quantum concurrence is defined as $$\mathcal{C}[\rho] = \mbox{max} \{\lambda_1 - \lambda_2 - \lambda_3 - \lambda_4, 0 \},$$ where $\lambda_1  >  \lambda_2  >  \lambda_3  >  \lambda_4$ are the eigenvalues of the operator $\sqrt{\sqrt{\rho}(\sigma^{(1)}_y \otimes \sigma^{(2)}_y) \rho^*(\sigma^{(1)}_y \otimes \sigma^{(2)}_y) \sqrt{\rho}}$. For pure states, the quantum concurrence is evaluated by
\begin{equation}
\mathcal{C}[\rho] = \sqrt{1 - a_1 ^2} = \sqrt{1 - a_2^2},
\end{equation}
with $a_{1,2}$ the modulus of the Bloch vectors associated to each subsystem, obtained via the Fano decomposition of the density operator
\begin{equation}
\rho = \frac{1}{4} \left[ \hat{I} +(\hat{\bm{\sigma}}^{(1)} \otimes \hat{I}^{(2)})\cdot \bm{a}_1+ (\hat{I}^{(1)} \otimes \hat{\bm{\sigma}}^{(2)})\cdot \bm{a}_2 + \displaystyle \sum_{i,j = 1} ^3 t_{ij} (\hat{\sigma}_i^{(1)} \otimes \hat{\sigma}_j^{(2)}) \right].
\end{equation}

\begin{figure}[!htb]
\includegraphics[width= 16 cm]{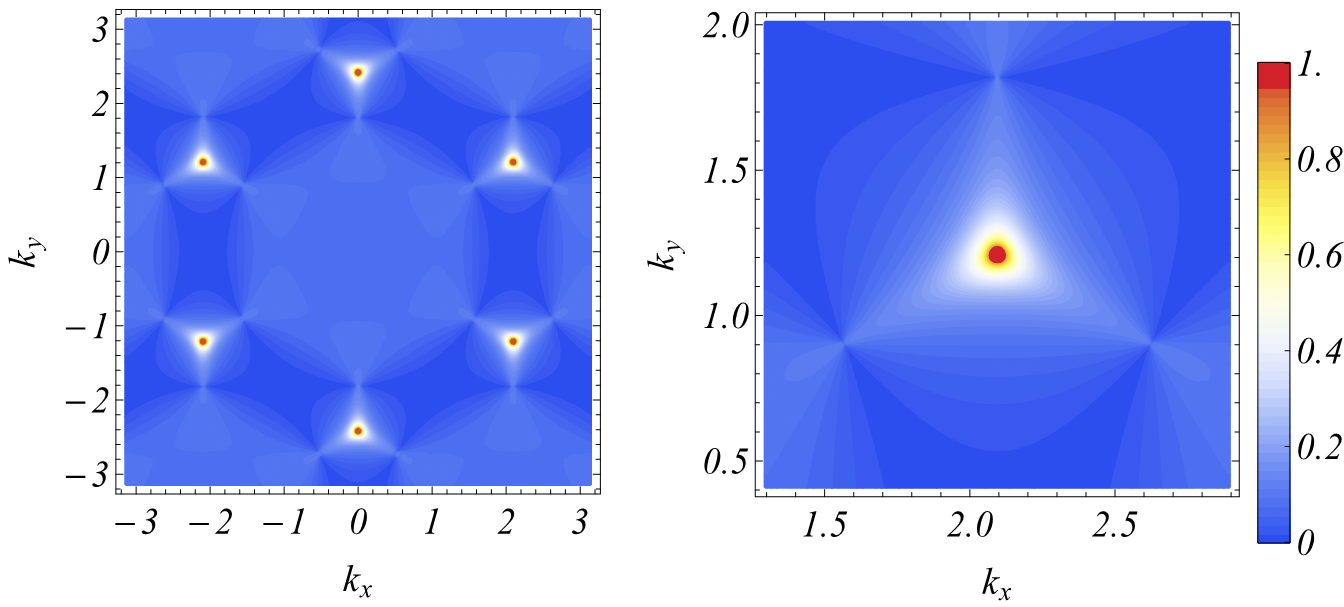}
\renewcommand{\baselinestretch}{1.0}
\caption{(Left) Concurrence in $\bm{k}$ space for $s=1$, $m=0$, $\Lambda = 0$ and the same set of parameters adopted in Fig.~\ref{energy}. (Right) When both bias voltage and mass parameters are missed, i.e., for the AB Hamiltonian Eq.~(\ref{ABHamiltonian}), the lattice-layer quantum correlations are maximized for states near the Dirac points, which depicts concurrence around $\bm{K}_+$.}
\label{concurrencem0l0}
\end{figure}

Once the tools for computing quantum entanglement and for describing the dynamics driven by the AB tight binding Hamiltonian are settled, the correlational properties of the eigenstates of (\ref{DiracHamiltonian}) can be computed. The density matrices generated by means of (\ref{ansatz}) have the Bloch vector $\bm{a}_2$ given by
\begin{eqnarray}
\label{BlochVector}
\bm{a}_2 &=& \mbox{Tr}[(\hat{I}^{(1)} \otimes \hat{\bm{\sigma}}^{(2)}) \rho_{n,s}] = \nonumber \\
&=&\frac{(-1)^n}{\vert \lambda_{n,s} \vert} \left[ \bm{W}  + \frac{(-1)^s}{\sqrt{g_2}}\left[(\bm{p} \cdot \bm{W}) \bm{p} + (\bm{W} \cdot \bm{\mathcal{E}}) \bm{\mathcal{E}} + M (M \bm{W} + \bm{p}\times \bm{\mathcal{E}}) \right] \right],
\end{eqnarray}
and, through graphene and Dirac parameter relations from (\ref{relations}), one can quantify the {\em lattice-layer} entanglement for the eigenstates described by $\rho_{n,s}$. Results do not depend on the quantum number $n$ and, therefore, the entanglement profile does not depend on the energy band to which the state belongs.
Figure \ref{concurrencem0l0} depicts the contour plots of concurrence as function of the $k_x$ and $k_y$ components of the wave vector $\bm{k}$ in the first Brillouin zone for $m=0$ and $\Lambda = 0$, i.e., when the dynamics is only driven by Eq.~(\ref{ABHamiltonian}). Entanglement is highly concentrated around the Dirac points. One notices that for $m = \Lambda = 0$ the wave function is singular at $\bm{K}_\pm$. At these points one has separable mixed states.

\begin{figure}[!htb]
\includegraphics[width= 17 cm]{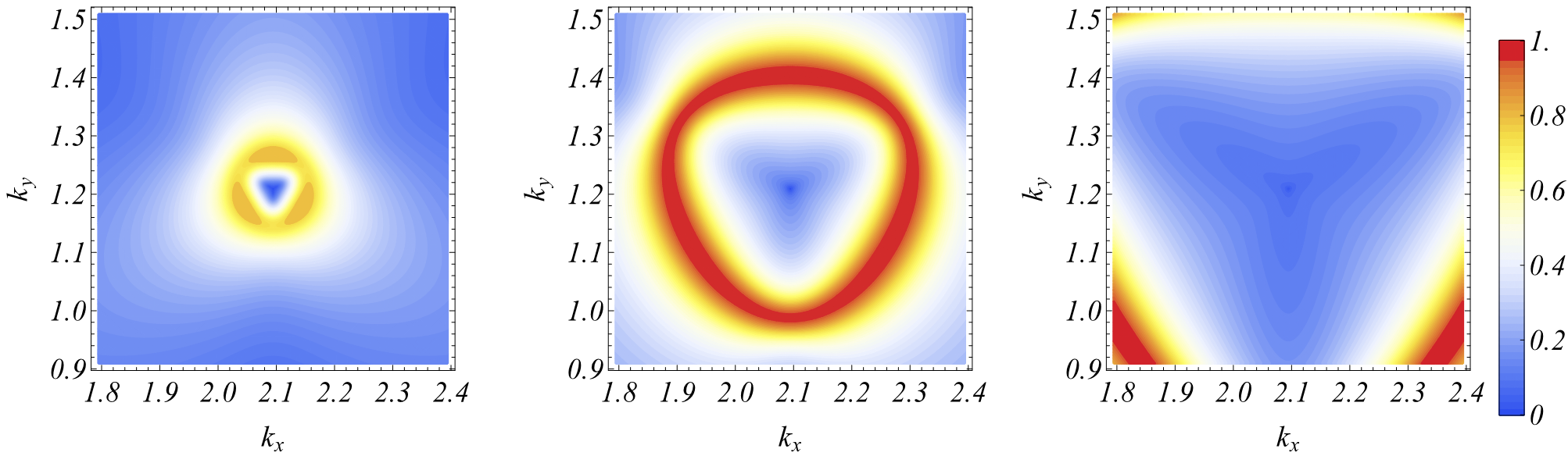}
\renewcommand{\baselinestretch}{1.0}
\caption{Concurrence in the $\bm{k}$ space for $s=1$ around $\bm{K}_+$, for the same set of parameters adopted in Fig.~\ref{concurrencem0l0}, with $m/t_\bot=0$ and $\Lambda/t_\bot = 1$ (first plot), $\Lambda/t_\bot = 5$ (second plot), and $\Lambda/t_\bot = 10$ (third plot). The inclusion of the bias voltage Hamiltonian (\ref{biasvoltageHamiltonian}) spreads lattice-layer entanglement around the $k$ space. Considering the angular coordinate on the $k_x-k_y$ plane, maximally entangled states correspond to the red region around the Dirac point.}
\label{concurrencelambda}
\end{figure}

Moreover, for $m \neq 0$ and/or $\Lambda \neq 0$, states with $\bm{k} = \bm{K}_\pm$ are separable for $s = 1$. For $s = 0$ states with Dirac momentum have concurrence given by
\begin{equation}
\mathcal{C}[\rho_{n,0}(\bm{K}_\pm)] = \frac{4 t_\bot^2}{4 t_\bot^2 + (2 m + \Lambda)^2}.
\end{equation}
The effects of a non vanishing $\Lambda$ contribution around the Dirac point $\bm{K}_+$ are depicted in Fig.~\ref{concurrencelambda}.
The bias voltage term spreads concurrence over the reciprocal space, distributing entanglement around the Dirac points. The larger the value of $\Lambda$, the more spread out is the distribution of entanglement in the $k$ space. Otherwise, when one considers $\Lambda/t_\bot \rightarrow \infty$, concurrence vanishes for all wave vectors.

\begin{figure}[!htb]
\includegraphics[width= 17 cm]{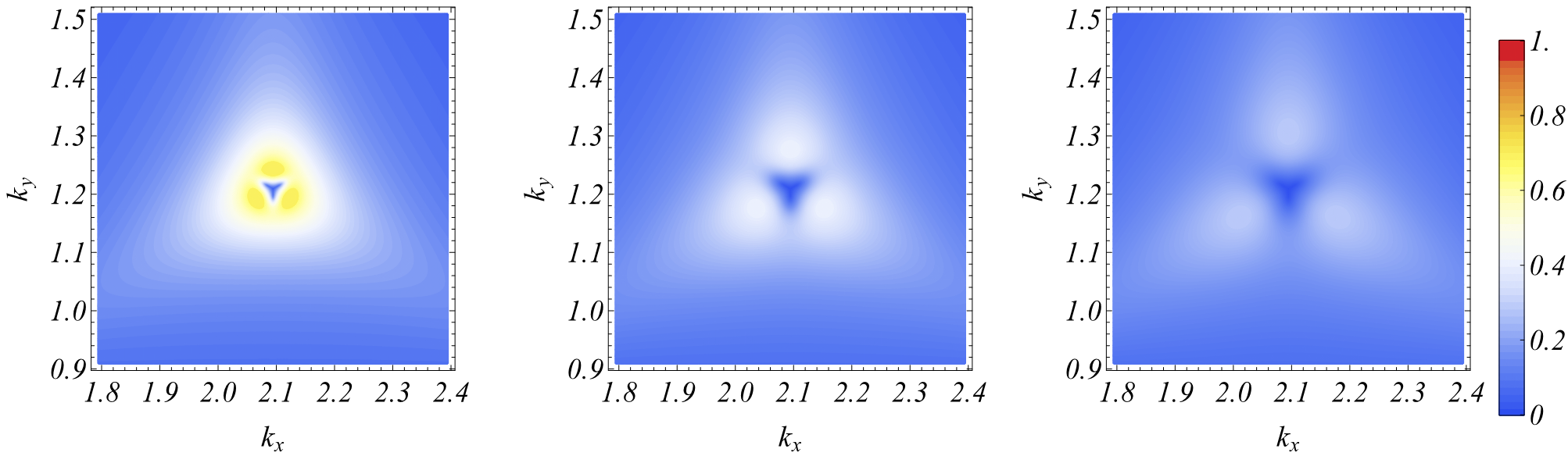}
\renewcommand{\baselinestretch}{1.0}
\caption{Concurrence in the $\bm{k}$ space for $s=1$ around $\bm{K}_+$, for the same set of parameters adopted in Fig.~\ref{concurrencem0l0}, for $\Lambda/t_\bot=0$, with $m/t_\bot = 0.1$ (first plot), $m/t_\bot = 0.5$ (second plot), and $m/t_\bot = 1$ (third plot). The corresponding mass term (\ref{massHamiltonian}) destroys the quantum entanglement. For $m/t_\bot \rightarrow \infty$, the entanglement vanishes in the whole first Brillouin zone.}
\label{concurrencemass}
\end{figure}

\begin{figure}[!htb]
\includegraphics[width= 17 cm]{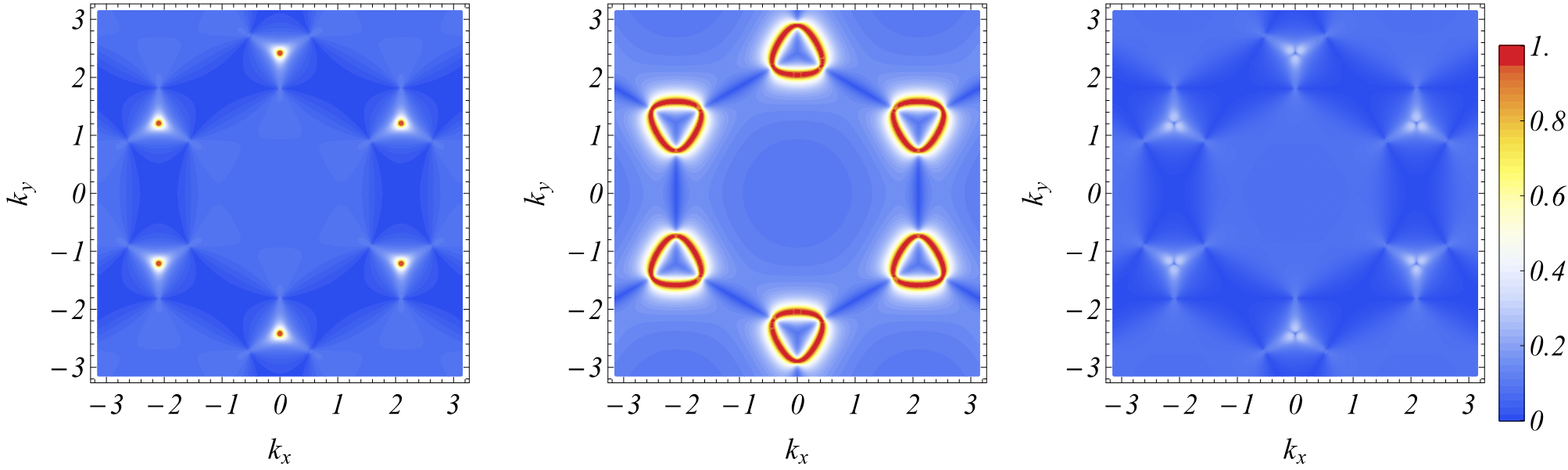}
\renewcommand{\baselinestretch}{1.0}
\caption{Entanglement profile comparison for $\bm{k}$ in the first Brillouin zone. The plots are for $(\Lambda/t_\bot, m/t_\bot) = (0,0)$ (first plot), $(\Lambda/t_\bot, m/t_\bot) = (10,0)$ (second plot), and $(\Lambda/t_\bot, m/t_\bot) = (0,1)$ (third plot), other parameters are fixed with the same values adopted in Fig.~\ref{concurrencem0l0}.}
\label{concurrencecomp}
\end{figure}

One also notices that the entanglement rapidly decreases as a function of $m/t_\bot$ when $\Lambda/t_\bot = 0$. Figure \ref{concurrencemass} depicts the density plot of concurrence in $\bm{k}$ space around $\bm{K}_+$ for the same set of parameters adopted in Fig.~\ref{concurrencem0l0}, with $\Lambda/t_\bot = 0$ and $m/t_\bot = 0.1$ (first plot), $m/t_\bot = 0.5$ (second plot), and $m/t_\bot = 1.0$ (third plot).
The mass parameter increasing generically suppresses the \textit{lattice-layer} entanglement, as depicted in Fig.~\ref{concurrencemass}.
Figure \ref{concurrencecomp} depicts quantum concurrence in the first Brillouin zone for $(\Lambda/t_\bot, m/t_\bot) = (0,0)$ (first plot), $(\Lambda/t_\bot, m/t_\bot) = (10,0)$ (second plot), and $(\Lambda/t_\bot, m/t_\bot) = (0,1)$ (third plot). While bias voltage spreads entanglement over the $\bm{k}$ space, the mass term suppresses it. States with momentum in the center of the first Brillouin zone (for low values of $k$) are separable.

\begin{figure}[!htb]
\includegraphics[width=16 cm]{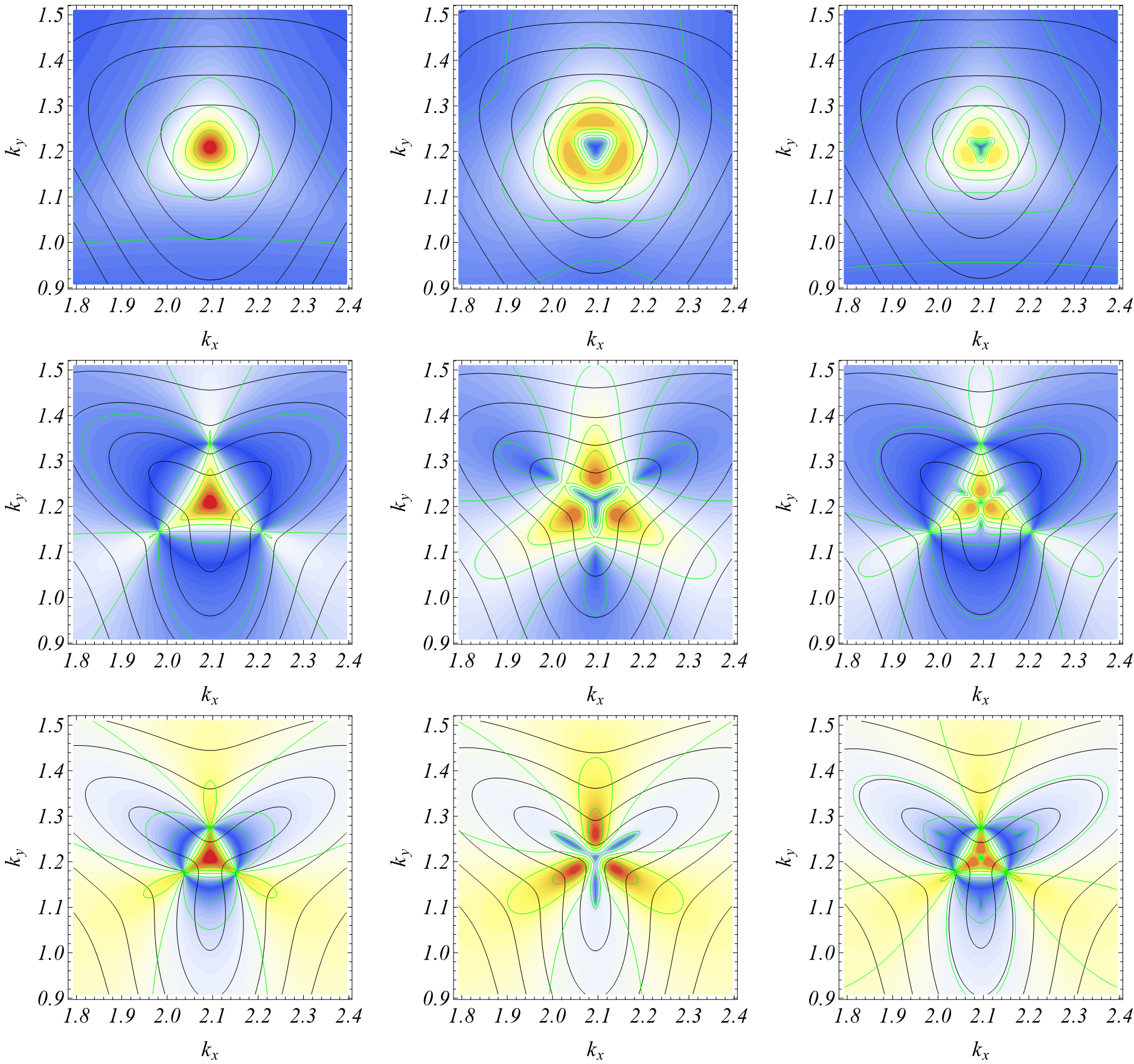}
\renewcommand{\baselinestretch}{1.0}
\caption{(Colors on line) Contour plot of the concurrence near a Dirac point for $t_3/t_\bot = 0.997$ (first row), $t_3/t_\bot = 5$ (second row) and $t_3/t_\bot = 10$ (third row), for $m/t_\bot = \Lambda/t_\bot = 0$ (first column), $m/t_\bot= 0$ and $\Lambda/t_\bot = 1$ (second column), and $m/t_\bot= 1$ and $\Lambda/t_\bot = 0$ (third column). Additional parameters are in correspondence with the previous plots. Black lines correspond to isoenergy lines while green lines correspond to isoconcurrence lines. Concurrence also exhibits a $\frac{2 \pi}{3}$ polar symmetry such that the distortion of the entanglement has a behavior similar to those exhibited by the eigenenergies due to the $t_3$ hopping parameter.}
\label{concurrenceenergy}
\end{figure}

The trigonal warping identified  on the energy spectrum of the AB Hamiltonian has also some implications onto the physical properties of the bilayer graphene \cite{graph03, graph04,Trig01, Trig02} and on the entanglement spectrum of its eigenstates \cite{Predin}. One thus may investigate the effects of the $t_3$ hopping parameter on the intrinsic {\em lattice-layer} entanglement. Figure \ref{concurrenceenergy} depicts concurrence density plots superimposed by the corresponding iso-energy lines (black lines) and by the iso-entanglement lines (green lines) for $t_3/t_\bot = 0.997$ (first row), $t_3/t_\bot = 5$ (second row) and $t_3/t_\bot = 10$ (third row), and with $m/t_\bot = \Lambda/t_\bot = 0$ (first column), $m/t_\bot = \Lambda/t_\bot = 0$ (first column), $m/t_\bot= 0$ and $\Lambda/t_\bot = 1$ (second column), and $m/t_\bot= 1$ and $\Lambda/t_\bot = 0$ (third column). The distortion exhibited by the isoenergy lines is also identified in the concurrence pattern, with similar angular symmetry profile, which is invariant under rotations of $\frac{2 \pi}{3}$. In fact, the isoentanglement lines have two symmetry patterns: the first one follows the isoenergy lines, and second one is rotated by and additional $2 \pi/3$ angle with respect to the isoenergy pattern. Again, after suppressing the $t_3$ contribution, the entanglement recovers its symmetry around the Dirac points. Otherwise, increasing $t_3/t_\bot$ values leads to an overall increasing entanglement.

It is worth mentioning that a more realistic description of graphene will include disorders, such as impurities and vacancies. Within the framework of the tight binding model, one can input effects of local impurities by including short-ranged potentials (such as the Coulomb potential) in the Dirac equation and considering scattering processes, from which transport properties can be derived \cite{Imp01, Imp02, Imp03, Imp04, graph01}. Moreover, disorders can nucleate localized states, which affects the density of states of the system \cite{graph01, Imp05} and also generates spin-orbit couplings \cite{Imp06}. The spherical wave scattering, describing charged impurities, can be evaluated in a framework similar to that used to compute \textit{spin-parity} entanglement under a barrier scattering \cite{Barrier}. The incident amplitudes will govern the entangling properties of both reflected and transmitted states, and one expect to observe features such as entanglement generation/destruction by the scattering process with the impurity. The lattice-layer entanglement in localized states induced by impurities can also be evaluated directly through the solutions of the Dirac equation with the corresponding external potential. In this case, entanglement will depend on the localization of the state, and features such as entanglement oscillation will be exhibited. The above features will be considered, from both graphene and more generic mathematical points of view, in subsequent work.

\section{Conclusions}

Since the techniques of preparing samples of graphene were devised, the theoretical and phenomenological description of this material has attracted great interest from several application perspectives \cite{graph01, graph02, graph03, graph04, graphteo}. In particular, the description of low energy electronic excitations of both monolayer and bilayer graphene by means of a suitable Dirac-like equation has introduced the perspective of observing and measuring quantum correlations in graphene.

In this paper we considered an extension of the relation between graphene and Dirac equation dynamics, related to the computation of intrinsic {\em lattice-layer} quantum entanglement. In a preliminary approach, the tight binding Hamiltonian for the bilayer Bernal stacking graphene was written as a Dirac Hamiltonian including pseudovector and tensor external fields, in order to have a map between graphene phenomenological parameters and Dirac mathematical variables.
The observation that the bispinors, solutions of the Dirac equation, are in general {\em parity-spin} entangled states, has been shown to support an easy method to calculate {\em lattice-layer} quantum entanglement for the tight binding Hamiltonian considered here.

By interpreting each graphene state as a two-qubit state, with one qubit associated with the lattice quantum number and the other one with the layer quantum number, the {\em lattice-layer} entanglement was quantified by means of the quantum concurrence, where the effects of bias voltage and mass terms were included. All the strategies for eigenstate building and entanglement computation were supported by algebraic properties of the Hamiltonian mapped into a $SU(2)\otimes SU(2)$ Dirac structure.

Our results show that, phenomenologically, when bias voltage and mass terms are absent, {\em lattice-layer} entanglement is highly concentrated around the Dirac points, in the corners of the first Brillouin zone. The wave functions of states with the wave vector in the Dirac points are singular, since at these points the energy eigenvalues are null, and one can identify separable mixed states. When the bias voltage is recovered, the entanglement spreads along $\bm{k}$ space, and is symmetrically distributed around the Dirac points. Otherwise, the mass term actively suppresses the quantum entanglement. For moderate values of mass, the eigenstates become weakly entangled for any wave vector. Moreover, in both cases, when bias voltage and mass terms are turned on, the states in the Dirac points are separable for the energy branch $s = 1$.

Finally, additional effects due to the trigonal warping onto the quantum concurrence were also quantified. It has been demonstrated that the distortion on the concurrence caused by the $t_3$ hopping is typically similar to the distortion of the isoenergy line. In fact, the entanglement exhibits the same $\frac{2 \pi}{3}$ polar symmetry around the Dirac points observed in the energy bands, with additional structures rotated with respect to the energy behavior.

To conclude, we point to the manipulation and measurement of the above discussed graphene properties \cite{graph01, graph02, Exp01, Exp02, Exp03, Manip, Manip02, Manip03}. Graphene systems usually include edge states \cite{Edge01, Edge02} and quantum dots states \cite{QuantumDot01, QuantumDot02, QuantumDot03} which can be manipulated and characterized by means of optical techniques and experimental processes based on measurement of electronic properties \cite{Exp02, Manip02, QuantumDot03}. Although no protocol for state tomography, i.e., a complete reconstruction of the density matrix throughout experimental measurement, has been proposed to date, the experimental measurement of quantum entanglement computed in this paper for single-particle excitations is an involving task which can be considered in the eventual construction of such protocols, since a complete characterization of intrinsic {\em lattice-layer} entanglement of single particle excitations in bilayer graphene has now been provided.

Although no protocol for direct single-particle state manipulation in graphene has been available, the literature on experimental characterization of graphene is vast. It also creates a challenging environment for future proposals of building quantum gates using the qubit assignment of Eq. (15). The use of monolayer graphene states as qubits was previously considered \cite{QuantumInfoGraph03, QuantumInfoGraph03}, as well as was implementing quantum gates using graphene with other physical systems \cite{QuantumInfoGraph01, QuantumInfoGraph02}. Nevertheless, the intrinsic entanglement can be used as a resource to improve quantum information protocols similar to those of quantum optics \cite{photon05}, and it can also be used to investigate features such as nonlocality \cite{Neut02}. Moreover, one might consider the extension of the protocol presented in \cite{Neut03} to map the intraparticle lattice-layer entanglement to interparticle entanglement, which might be more suitable for implementation of quantum gates as well as other quantum information protocols.

Regarding future extensions related to the present formalism one might include different layer arrangements, such as AA stacking and twisted graphene \cite{graph03}, as well as wave packet and other localization effects \cite{Wave01, Wave02, Wave03}.

{\em Acknowledgments - The work of A. E. B. is supported by the Brazilian Agencies FAPESP (Grant No. 15/05903-4) and CNPq (Grant No. 300809/2013-1). The work of V. A. S. V. B. is supported by the Brazilian Agency CNPq (Grant No. 140900/2014-4).}

\section*{Appendix}

The total Hamiltonian from Eq.~(\ref{Htotal}) reads
\begin{eqnarray}
\label{Htotal}
\hat{\mathcal{H}} = \hat{\mathcal{H}}_{AB} + \hat{\mathcal{H}}_m + \hat{\mathcal{H}}_{\Lambda} = \left[ \, \begin{array}{cccc}
\frac{\Lambda}{2} + m & - t \Gamma(\bm{k}) & 0 & - t_3 \Gamma^*(\bm{k})\\
- t \Gamma^*(\bm{k}) & \frac{\Lambda}{2} - m & t_\bot & 0\\
0 & t_\bot & - \frac{\Lambda}{2} + m & -t\Gamma (\bm{k})\\
- t_3 \Gamma(\bm{k}) & 0 & - t \Gamma^*(\bm{k}) & -\frac{\Lambda}{2} - m
\end{array} \right].
\end{eqnarray}
such that each term of $\hat{\mathcal{H}}_{AB}$ can be written in terms of the decomposition
\begin{eqnarray}
\label{rel01}
\left[ \, \begin{array}{cccc}
0 & 0 & 0 & 0\\
0 & 0 & t_\bot & 0\\
0 & t_\bot & 0& 0\\
0 & 0 & 0 & 0
\end{array} \right] &=& \frac{t_\bot}{2}\left( \hat{\alpha}_x - i \hat{\gamma}_y\right),  \\
- t \left[ \, \begin{array}{cccc}
0 & \Gamma(\bm{k}) & 0 & 0\\
 \Gamma^*(\bm{k}) & 0 & 0 & 0\\
0 & 0 & 0& \Gamma (\bm{k})\\
0 & 0 & \Gamma^*(\bm{k}) & 0
\end{array} \right] &=& - t \left\{\mbox{Re}[\Gamma (\bm{k})] \hat{\gamma}_5 \hat{\alpha}_x -\mbox{Im}[\Gamma(\bm{k})] \hat{\gamma}_5 \hat{\alpha}_y\right\},
\end{eqnarray}
\begin{eqnarray}
\label{rel02}
- t_3 \left[ \, \begin{array}{cccc}
0 & 0 & 0 & \Gamma^*(\bm{k})\\
0 & 0 & 0 & 0\\
0 & 0 & 0& 0\\
\Gamma(\bm{k}) & 0 & 0 & 0
\end{array} \right] = -\frac{t_3}{2}\left\{ \mbox{Re}[\Gamma (\bm{k})](\hat{\alpha}_x + i\,\hat{\gamma}_y) + \mbox{Im}[\Gamma(\bm{k})]( \hat{\alpha}_y - i\, \hat{\gamma}_x)\right\},
\end{eqnarray}
and one has identified
\begin{eqnarray}
\label{rel03}
\hat{\mathcal{H}}_m = m \hat{\gamma}_5 \hat{\alpha}_z, \hspace{1 cm} \hat{\mathcal{H}}_{\Lambda} = \frac{\Lambda}{2}\hat{\beta},
\end{eqnarray}
for mass and bias voltage terms.

\end{document}